\documentclass[conference]{IEEEtran}% Add the compsocconf option for 
\ifCLASSINFOpdf
\else
\fi

\usepackage{amsmath}
\usepackage{algorithm}
\usepackage{caption}
\usepackage[noend]{algpseudocode}
\DeclareCaptionFormat{myformat}{#3}
\usepackage{graphicx}
\begin{document}
%\title{Real Time Twitter Sentiment Analysis Using Apache Spark}
\title{Real-time Text Analytics Pipeline Using Open-source Big Data Tools}


\author{\IEEEauthorblockN{Hassan Nazeer, Waheed Iqbal, Fawaz Bokhari, \\Faisal Bukhari}
\IEEEauthorblockA{P.U.C.I.T University of the Punjab, \\ Lahore, Pakistan.}
\and
\IEEEauthorblockN{Shuja Ur Rehman Baig}
\IEEEauthorblockA{Technical University of Catalonia and \\Barcelona Supercomputing Center,\\
Barcelona, Spain.}
}

\maketitle

\begin{abstract}
Real-time text processing systems are required in many domains to quickly identify patterns, trends, sentiments, and insights. Nowadays, social networks, e-commerce stores, blogs, scientific experiments, and server logs are main sources generating huge text data. However, to process huge text data in real time requires building a data processing pipeline. The main challenge in building such pipeline is to minimize latency to process high-throughput data. In this paper, we explain and evaluate our proposed real-time text processing pipeline using open-source big data tools which minimize the latency to process data streams. Our proposed data processing pipeline is based on Apache Kafka for data ingestion, Apache Spark for in-memory data processing, Apache Cassandra for storing processed results, and D3 JavaScript library for visualization. We evaluate the effectiveness of the proposed pipeline under varying deployment scenarios to perform sentiment analysis using Twitter dataset. Our experimental evaluations show less than a minute latency to process $466,700$ Tweets in $10.7$ minutes when three virtual machines allocated to the proposed pipeline.
\end{abstract}

\begin{IEEEkeywords}
Big Data Processing; Apache Spark; Apache Kafka; Real-time Text Processing; Sentiment Analysis.

\end{IEEEkeywords}

\IEEEpeerreviewmaketitle

\section{Introduction}
Today real-time analytics for text data on large-scale has become important for many business needs. Comparing to traditional data warehouse applications, the real-time analytic are data intensive in nature and require to capture and process the data efficiently. However, collecting and processing such a large-scale data had introduced new challenges in terms of storage as well as processing time.

A typical real-time data processing of large-scale data requires building a distributed data pipeline for capturing, processing, storing, and analyzing the data efficiently. The real-time processing system should be capable of capturing high rate data from various streaming sources, process the data near real time, and store data into a persistent database. However, the data processing system should provide minimum latency to process a high throughput data in real time which is a challenging job.

Google introduced the Map Reduce paradigm \cite{google-MapReduce} for parallel and distributed execution of an application over the commodity cluster. Several systems had implemented Map Reduce paradigm for parallel and distributed processing of batch data on multiple machines. Apache Hadoop is one of the most well-known implementations of it. However, recently Apache Spark \cite{spark} gets more attraction mainly due to the extended capabilities of Hadoop echo system and allows to process real-time streaming data. 

Distributed messaging systems are mainly used for data ingestion for real-time processing. Distributed messaging systems work on a publish-subscribe model where all incoming messages are broadcasted to all subscribed consumers. RabbitMQ \cite{rabbitmq}, Apache Kafka \cite{kafka}, and ActiveMQ \cite{activemq} are famous open-source distributed messaging systems that are widely used as data ingestion systems for large-scale data processing. 

Distributed storage systems are used to store big data. Nowadays, NoSQL-based systems are famous for providing distributed storage and retrieval mechanism with good performance and scalability. MongoDB \cite{mongodb}, Cassandra \cite{cassandra}, HBase \cite{hbase}, and Redis \cite{redis} are widely used NoSQL-based systems. 

In this paper, we have proposed and evaluated a real-time processing pipeline using the open-source tools that can capture a large amount of data from various data sources, process, store, and analyze the large-scale data efficiently. Our proposed system uses Apache Kafka \cite{kafka} as data ingestion system, Apache Spark \cite{spark} as a real-time data processing system, Apache Cassandra for persistent distributed storage, and D3 \cite{d3} JavaScript library for visualization.

In a traditional uses case, Apache Kafka accepts incoming data and sends it to Apache Kafka broker at a very rapid rate. Then Apache Spark consumes the data and performs predictive analytics using Spark's MLib module \cite{mlib}. Finally, we use Apache Cassandra connector components (3rd party) to store the data in Cassandra. We developed a simple visualization component to analyze the results using NodeJS application. 

To evaluate the proposed system, we developed a sentiment analysis application using Twitter data. We acquired a large set of Tweets using  Twitter streaming API. Our proposed sentiment analysis application processes each tweet and classifies it either positive or negative sentiment and store it for analytics purposes. The application deployed on the proposed pipeline, the Twitter data is streamed to Kafka which makes it available for Spark that performs the classification and store results into Cassandra. Then we use the visualization component to analyze the different sentiment trends.   

We performed several experiments to profile the system throughput, performance, and latency of the application under different resource allocations.  All of our experiments are performed on OpenNebula \cite{opennebula} private cloud that was established using commodity hardware. We measure the performance and latency in terms of execution time that application takes to process and store data into database, and throughput in terms of messages that are processed by the application.  We obtained a maximum of 41 seconds latency to process 466,700 Tweets in 10.6 minutes. Our experimental evaluations show that as we increase the number of resources to Spark, the time taken by the application to process the Tweets reduces which in turn increases the performance and reduces the latency of application. 
%i doubt on this lets confirm from the experiment
%However the throughput has no significant effect as we change number of machines for application.  

In the rest of this paper, we discuss the related work, system components to build the text analytics pipeline, experimental design, and experimental results in detail. 

\section{Related Work}
In recent years, there have been several contributions towards building real-time processing systems. For example Perera et al. \cite{Perera:2015:SPR:2675743.2774214} describe common real-time analytics use cases with implementation details. Liu et al. \cite{Liu:2014:SRP:2628194.2628251} present an overview of the open-source technologies that support big data processing in a real-time. However, these contributions do not address building a complete pipeline to process the big data in real time. 

Twitter data is used to build various real-time processing systems. For example,  authors in \cite{shoro2015big} use Apache Spark streaming to process Tweets posted on Twitter in real time. The authors collected the data through Twitter streaming API and process it through Spark streaming. Their objective was to find top 10 words over last 10 minutes, top 10 languages over last 10 minutes and occurrence of the particular word found in Tweets over the specific time period. Another contribution using Twitter data has been published in \cite{KhanMinhas:2015:IFI:2837185.2843853} which clusters Tweets in real-time, adjusts new Tweets into existing clusters and provides visualization of clusters that help in identifying latent topics and sub-topics within the Tweets. Jose et al. \cite{jose2016prediction} implements a real-time Twitter sentiment analyzer using classifier ensemble approach. In \cite{chamansingh2016efficient}, the authors have provided a sentiment analysis algorithm for Twitter feeds. The focus of the authors was to determine an efficient sentiment classifier of real-time Twitter feed. 

Authors in \cite{Maynard:2015:RSM:2786451.2786500} proposed an open-source framework for analytics of social media using semantic annotation and linked open data. In another contribution, authors in \cite{Grant:2012:MQS:2396761.2398746} presented a real-time declarative query processing system over multiple data sources with both structured and text information. Another paper \cite{nuesch2014real} presents a system to process sensor data in Water Distribution Network (WDN) using Apache Spark to detect the anomaly in WDNs using various statistical techniques. 

To the best of our knowledge, there is no work exist that build and evaluate a complete data analytic pipeline using open-source big data tools and technologies. In this paper, our work focuses on building a real-time text processing pipeline with minimal latency using open-source big data tools. We evaluate the performance of the proposed system using various deployment scenarios for a sample sentiment analysis applications using Twitter data. 

%Han et al. \cite{han2016topic} present Chinese semantic rules and combine the sentiment based and similarity based method to extract sentiment words and the results are revised again by emoticons.

%In \cite{solaimani2014statistical} anomaly detection in VMWare based data center has been performed by using statistical techniques in Apache spark in order to monitor VM based data center. They collect each machine performance metric data such as cpu usage, memory usage and send it to Apache spark streaming to detect anomaly.

%The Market basket Analysis \cite{woo2011market} measures the performance of Apache spark by performing Market Basket analysis using Apriori Algorithm in Apache spark with different data sets to measure the impact of number of nodes of cluster on spark streaming performance. 

\section{Real-time Text Analytics Pipeline}
%Waheed the heading of this section can be Our Proposed Text Analytics Pipeline System or something like that. what you say?
The real-time processing pipeline presented in this paper comprises the following five main components: 

\begin{enumerate}
\item Distributed Data Ingestion System
\item Distributed and Parallel Processing System
\item Distributed Database System
\item Multiple Streaming Sources
\item Visualization Component
\end{enumerate} 
In the rest of this section, we briefly explain each of these components in turn.
\subsection{Distributed Data Ingestion System}
In a real-time processing system, the purpose of data ingestion system is to collect and process the data for later use or database storage from multiple streaming sources. It is a major challenge for data ingestion system to steadily ingest the feeds of data from multiple streaming sources. As the data is ingested in a real-time system, it is processed by some real-time processing engine to accomplish the specific task. There exists multiple tools that can be used as data ingestion system in a real-time processing system such as Apache Kafka, Apache Flume \cite{flume} and RabbitMQ \cite{Rabitmq}. We used Apache Kafka as a data ingestion system for a real time processing system. Apache Kafka is a distributed high throughput publish-subscribe messaging system. In Apache  Kafka, the multiple producers publish the message on a topic and multiple consumers subscribe to that topic can also consume the messages.The major terms used in Apache Kafka are:
\begin{itemize}
\item \textbf{Topic:} Topic is a category that maintains the number of messages.
\item \textbf{Producer:} The application which produces the messages on a topic using Kafka API.
\item \textbf{Consumer:} The application which reads messages from a topic using Kafka API.
\item \textbf{Broker:} Broker is a Kafka cluster which consists of multiple nodes.
\end{itemize}
Apache Kakfa internally use Apache Zookeeper\cite{zookeeper} to maintain several activities across the Kafka cluster. 

\subsection{Distributed and Parallel Processing System}
The immense growth of data generated from the variety of data sources has changed the way to process and store the data. There are some scenarios, where the size of data is too large to process it on a single system, and we need parallel and distributed execution of data on multiple systems. There exist multiple tools that allow us to write parallel and distributed applications such as Apache Hadoop, Apache Spark, and Apache Storm\cite{storm}. For our proposed pipeline, we have used Apache Spark as a distributed and parallel processing system for the real time processing system. Apache Spark is in-memory cluster computing framework that was initially developed to run iterative algorithms based on machine learning. Apache Spark was developed by the University of Berkeley to overcome the limitations of Apache Hadoop\cite{hadoop}. Apache Spark stores all intermediate results in memory rather than storing them on a disk, which makes it 100 times faster than the Apache Hadoop. Figure \ref{fig1:Big_Data} shows the execution architecture of Apache Spark application:

\begin{figure}[!htb]
\begin{center}
  \includegraphics[width=\linewidth,height=15cm,keepaspectratio]{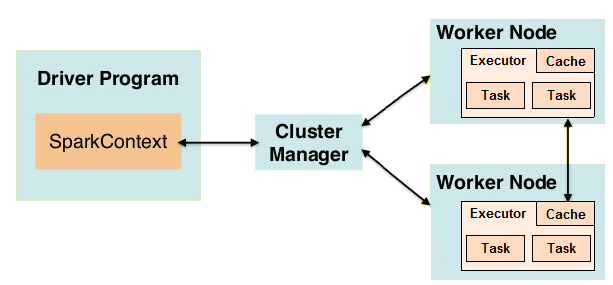}
  \caption{Execution Flow of Apache Spark application.}
  \label{fig1:Big_Data}
  \end{center}
\end{figure}

The driver process initiates multiple worker processes, each of which reads input data from HDFS \cite{hdfs} or
other file system and stores computational results in a memory for iterative machine learning algorithm. The abstract data type for distributed and parallel computing for Apache park is resilient distributed datasets (RDD's).

The resilient distributed datasets (RDD's) \cite{rdd} are partitioned collection of records that cannot be changed once created (immutable). RDDs can be created from input datasets or from applying some operation on existing RDDs. There are two different types of operations we can perform on RDDs, one is called transformations and other is called actions. The transformations when applying on RDDs create new RDDs. The example of transformations is map, group by key and filter etc. The action when applying on RDDs generates the aggregate value. The example of action are reduce, count, and take etc. The RDDs can be cached in memory on worker nodes in case of iterative machine learning algorithm so that the computed value from the previous iteration cannot be recomputed again.
The Apache Spark program is also based on map reduce paradigm as Apache Hadoop program but it has different processing mode. Apache Hadoop can only run in batch processing mode while Apache Spark can run in both real-time as well as batch processing mode.

\subsection{Distributed Database System} 
The NoSQL databases\cite{nosql} system are non-relational, distributed database system that allows the ad-hoc and fast analysis of high-velocity data with disparate data types. In fact, NoSQL databases system become an alternative of traditional RDMS system with keeping scalability, high availability, and fault tolerance as major key factors. There are a number of NoSQL databases available in the market such as Apache Hbase\cite{hbase}, Apache Cassandra and Mongodb\cite{mongodb}. We use Apache Cassandra as NoSQL distributed database for the real-time system.
Apache Cassandra is fully distributed decentralized NoSQL database that provides high availability of data, ease of operations and easy distribution of data across multiple data centers
builds on the top of the cluster. Apache Cassandra was initially developed at Facebook for solving slow search in the inbox and later on was converted to open source under Apache in the year 2010.
The following are the major terms used in Apache Cassandra:
\begin{itemize}
\item \textbf{Node:} An individual machine where data is stored.

\item \textbf{Data center:} It is a collection of connected nodes.

\item \textbf{Cluster:} A Cluster contains one or more data centers.

\item \textbf{Column:} A column is a tuple of name, value and time stamp.

\item \textbf{Super Column:} A tuple of name and value in which value is another column.

\item \textbf{Column Family:} A column family is a table similar to table in RDMS that contain infinite number of rows.

\item \textbf{Keyspace:} A Keyspace is a container that maintains multiple column families. There are two settings related to Keyspace. One is the key replication factor, which is the number of nodes in a cluster copying the same data for providing high availability of data, in the case of any node failure. The other is a replica placement strategy which is placement strategy of the replica.
\end{itemize}
Apache Cassandra architecture makes it easy, scalable and highly available database. Instead of traditional master-slave architecture, Apache Cassandra uses peer to peer distributed architecture, in which each node of the cluster is identical to other nodes in a cluster. The following figure shows the cluster or ring of four Apache Cassandra nodes:
\begin{figure}[!htb]
\begin{center}
  \includegraphics[width=\linewidth,height=4cm,keepaspectratio]{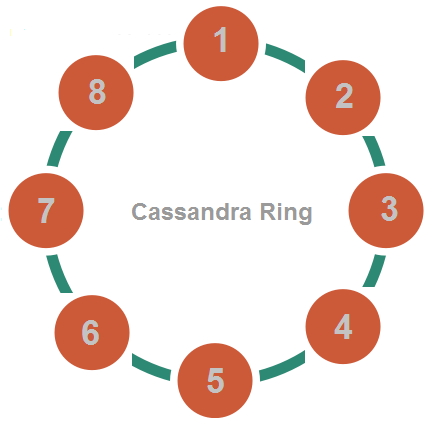}
  \caption{Apache Cassandra Ring.}
  \label{fig 1: Apache Cassandra Ring}
  \end{center}
\end{figure}

On the write operation, initially, each write is captured by commit log (mechanism of data recovery in Cassandra) which then write to an in-memory structure which is called Mem-Table. When Mem-Table is full, then data is written to a file called SS-table. All write data is automatically distributed and partitioned across all nodes in Apache Cassandra cluster. Apache Cassandra consolidates the SS-table to dismiss unnecessary data from time to time.

During a read operation, Apache Cassandra gets Mem-table value and consult bloom filter (algorithms for testing the membership of the element in a set) to find out the SS-table that contains requested data.

In our proposed pipeline, there are multiple Apache Kafka producer applications that send Tweets to a topic in an Apache Kafka broker, the Apache Spark streaming application reads the Tweets from the topic in Apache Kafka broker, classified Tweets, and store them in Apache Cassandra. Figure \ref{rows-cassandra} shows the 10 records in an Apache Cassandra column family (table):

\begin{figure}
\begin{center}
  \includegraphics[width=\linewidth,height=60cm,keepaspectratio]{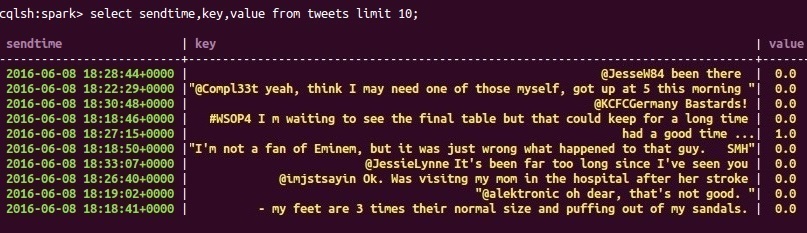}
  \caption{Records in Cassandra column family.}
  \label{rows-cassandra}
  \end{center}  
\end{figure} 

\subsection{Multiple Streaming Sources}
The streaming sources are data source points that push data into a real-time processing system.These sources can be sensors on social media websites or monitoring daemons. The role of a streaming source is to capture live data and push it to real-time processing for analysis and decision making. We use Twitter as a streaming source that pushes the live Tweets into real-time processing system to perform the sentiment analysis. Twitter provides a streaming API, that allows us to capture the live Tweets on Twitter. The data rate provided by Twitter streaming API for a free account is very low, it hardly contains three to four Tweets per seconds.To stimulate high data rate into our real-time processing system, first we create java application (Tweets Collector) that use Twitter Streaming API to get and store live Tweets into a file system and then another java application (Tweet Producer) that send Tweets from file system to Apache Kafka server at very rapid rate, approximately three hundreds Tweets per second.

\subsection{Visualization Component}
The visualization part of platform displays the real-time dashboard based on a real-time processed data, which helps in both decision making and visualization purposes. For, Twitter Sentiment Analysis Application, we created a web application in node.js technologies that utilize d3 JavaScript library to display graphs of some popular keyword found in positive and negative Tweets. The web application also allows users to search a particular keyword and finds its occurrence in positive and negative Tweets. The web application connects to Cassandra database, fetched the top 200 rows from the table, find the occurrence of each word in all Tweets and display a maximum of top 10 keywords that it has found in both positive and negative Tweets using d3 graph. 

Figure 4 shows the overall architecture of our proposed real-time processing pipeline.

\begin{figure}
 \begin{center}
  \includegraphics[width=\linewidth,height=22cm,keepaspectratio]{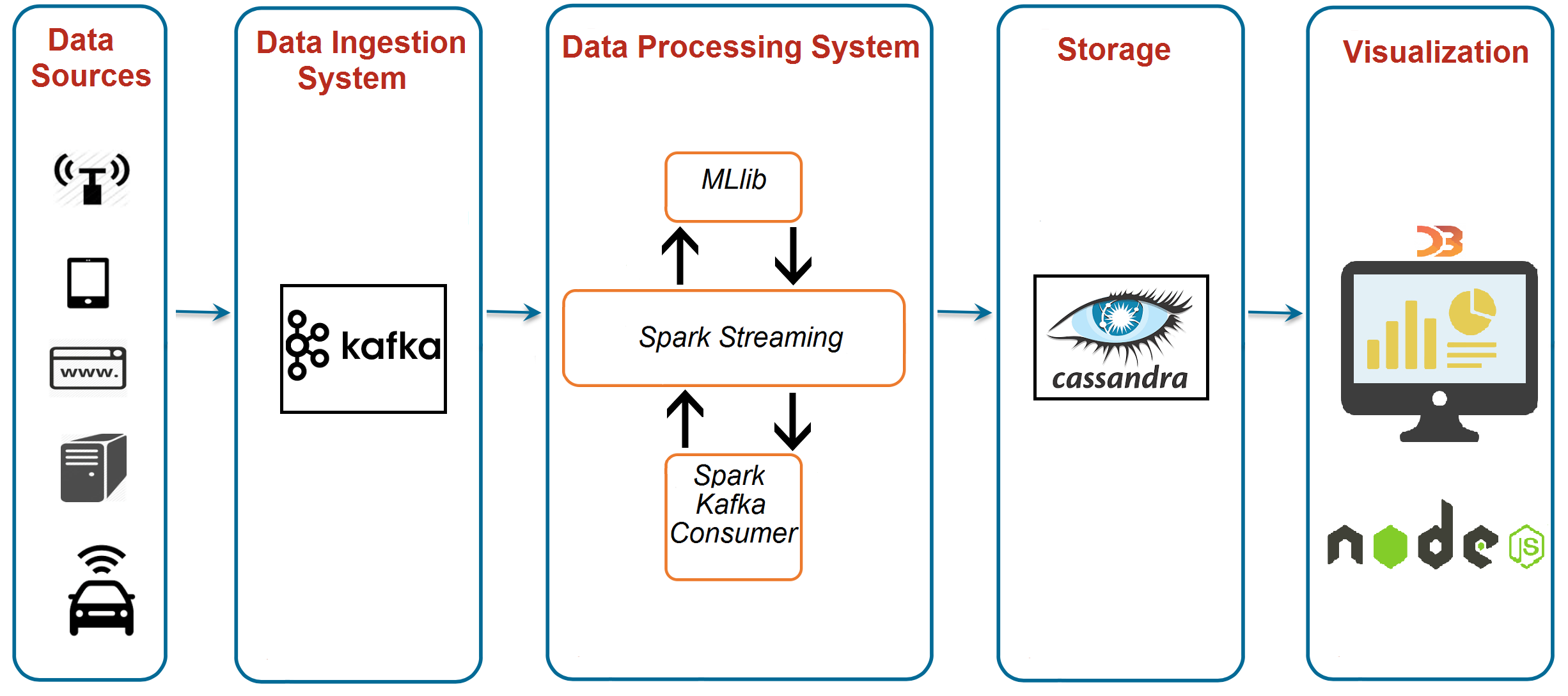}
  \caption{Proposed data analytics pipeline architecture.}  
  \end{center}
  \label{fig:archit}
\end{figure} 

\section{Experiment Design}
In this section, we describe testbed infrastructure, benchmark application, synthetic workload generation method, and experiments performed to evaluate the performance of our proposed system for real-time sentiment analysis. 

\subsection{Testbed Infrastructure}
To deploy Apache Spark, Apache Kafka, and Apache Cassandra we developed a small testbed private cloud using OpenNebula \cite{nebula} and commodity hardware. We used four homogeneous physical machines containing 3.6 GHz Core i7, 15 GB physical memory, and 1 TB hard disk. We have used a virtual machine template containing 2 vCPUS and 4 GiB physical memory to spawn a different number of virtual machines to conduct the experiments. 

\subsection{Benchmark Application: Sentiment Analysis using Tweets}

To evaluate the proposed text analytics pipeline, we built a Twitter sentiment analysis application. In Twitter, users can post 140 characters long text also called tweet which most reflect people opinions, discussion or products, political views, and news. Therefore, Twitter is a useful platform where we can find the trending topics, opinions, and discussion about the current affairs.We used Naive Bayes classifier \cite{naive_bayes} to build a sentiment analysis model. Naive Bayes Classifier uses probabilistic model to classify the text. In Naive Bayes Classifier, the label classes are already known and probabilistic model builds with help of training dataset to classify the new text whose class is not known. We pre-process the Tweets to remove irrelevant words such as stop words, URLs, numbers (date or time), and white space in both training and testing data sets, then we use Apache Spark's MLib implementation of Naive Bayes classifier \cite{mlib-naive} to train the model using 45,000 Tweets. When Spark starts, it trains the sentiment model and uses it to predict the sentiments of the unseen Tweets. 

\subsection{Workload Generation}
Twitter streaming API provides a limited number of Tweets per minute, therefore, we gathered a large number of Tweets offline and reply them in a fast rate to evaluate the performance of our proposed real-time system. We developed a Java application that replies the collected Tweets for a fixed number of Tweets per seconds to a specific Kafka topic. Spark streaming module consumes the Tweets from the topic. We run the workload generator using two machines to avoid any bottleneck on the workload generation. The workload generator replays the Tweets from a maximum of $2,020$ Tweets per second to a minimum $385$ requests per seconds.

%We created Apache Kafka producer application to push the data into realtime processing system. The Apache Kafka producer application push the Tweets into apache kafka to topic which apache spark streaming kafka consumer is subscribed.Due to limited rate of twitter streaming api for free account, we created the java application that uses twitter streaming api to collect the Tweets from twitter and save them into file system. The apache kafka producer application reads the Tweets from file system and push Tweets into apache kafka topic. 
%\\We executed the apache kafka producer application on seperate virtual machine with multiple instances each of which send Tweets from different part of file.The apache kafka producer application send Tweets to apache broker for 10 minutes.
%\\Each apache kafka producer application Instance takes rank and total number of instance as input  parameter, where rank is rank of current Instance and total number of instance are number of instances that should be launched parallel.
\subsection{Experiment Details}
We conducted three different experiments to evaluate the performance of the proposed system. Table \ref{table:summary-experiment} summarizes the experiments. In Experiment 1, we deployed Apache Kafka, Apache Spark, and Cassandra on a single virtual machine. In Experiment 2, we deployed a cluster of Apache Spark, Kafka, and Cassandra using two virtual machines. Each component of the pipeline is running as a cluster hosted on both virtual machines. In Experiment 3, we deployed a cluster of Apache Spark, Kafka, and Cassandra using three virtual machines. Each component of the pipeline is running as a cluster hosted on all three virtual machines. For all three experiments, we generated the same workload and profile system throughput and latency. 
\begin{table}[h!]
\begin{center}
\caption{Summary of Experiments.}
\begin{tabular}{p{2.5cm}p{5.2cm}}
  \hline
  \textbf{\textbf{Experiment} } & \textbf{Description}\\
  \hline
  \rmfamily 1. Simple Deployment & \rmfamily  Deployed Apache Spark, Apache Kafka and Apache Cassandra on a single virtual machine. \\ \hline
  
  \rmfamily 2. Distributed deployment with 2 instances & \rmfamily  Deployed Apache Spark, Apache Kafka and Apache Cassandra as a cluster using two virtual machines.  \\ \hline
  
  \rmfamily 3. Distributed deployment with 3 instances  &\rmfamily Deployed Apache Spark, Apache Kafka and Apache Cassandra as a cluster using three virtual machines.  \\ 
  \hline
  \label{table:summary-experiment}
\end{tabular}
\end{center}

\end{table}
%%%fig 1: Big Data
\section{Experiment Results}      
\subsection{Experiment 1: Simple Deployment}
\begin{figure}[!htb]
\begin{center}
   \includegraphics[width=\linewidth,height=5cm,keepaspectratio]{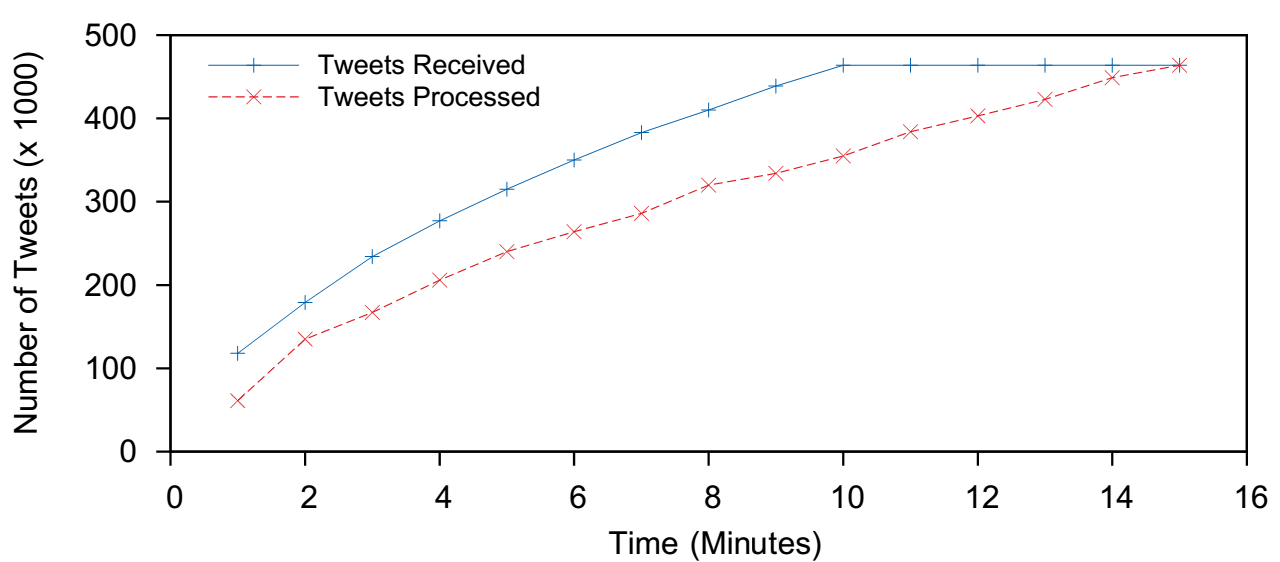}
  \caption{Number of received and process Tweets during the Experiment 1.}
  \label{fig:exp1}
  \end{center}
  %\label{fig:exp1}
\end{figure}

This section described the results obtained from the Experiment 1. The Apache Kafka producer application executed for 10 minutes to send Tweets to the system.  Figure \ref{fig:exp1} shows the aggregated Tweets received and processed during the Experiment 1. The total Tweets send by Apache Kafka producer application are 464,200. The Apache streaming application took 15 minutes to process all the Tweets. The overall latency in this experiment is 5 minutes. We consider this experiment as a baseline to calculate speedup of processing time in Experiment 2 and 3. 

\subsection{Experiment 2: Distributed Deployment with 2 Instances}
\begin{figure}[!htb]
\begin{center}
  \includegraphics[width=\linewidth,height=5cm,keepaspectratio]{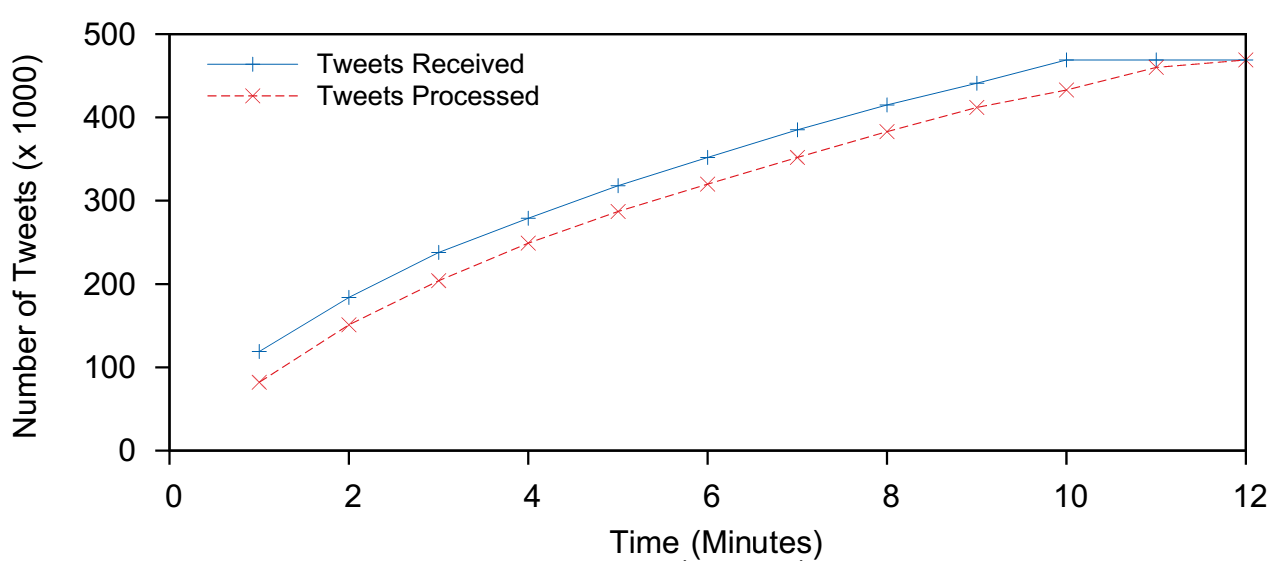}
  \caption{Number of received and process Tweets during the Experiment 2.}
  \label{fig:exp2}
   \end{center}  
\end{figure}

This section describes the results obtained from the Experiment 2.The Apache Kafka producer application executed for 10 minutes to send Tweets to system. Figure \ref{fig:exp2} shows the aggregated Tweets received and processed during the Experiment 2. The total Tweets send by Apache Kafka producer application are 468,600. The Apache streaming application took 11.5 minutes to process all the Tweets. The overall latency time in this experiment is 1.5 minutes. The speedup obtained to process the Tweets is 130\% comparing to Experiment 1. 

\subsection{Experiment 3: Distributed Deployment with 3 Instances}
This section describes the results obtained from the Experiment 3.The Apache Kafka producer application executed for 10 minutes to send Tweets to system.  Figure \ref{fig:exp2} shows the aggregated Tweets received and processed during the Experiment 3. The total Tweets send by Apache Kafka producer application are 466,700. The Apache streaming application took 10.7 minutes to process all the Tweets. The overall latency time in this experiment is 0.7 minutes. The speedup obtained to process the Tweets is 140\% comparing to Experiment 1.
\begin{figure}[!htb]
\begin{center}
  \includegraphics[width=\linewidth,height=5cm,keepaspectratio]{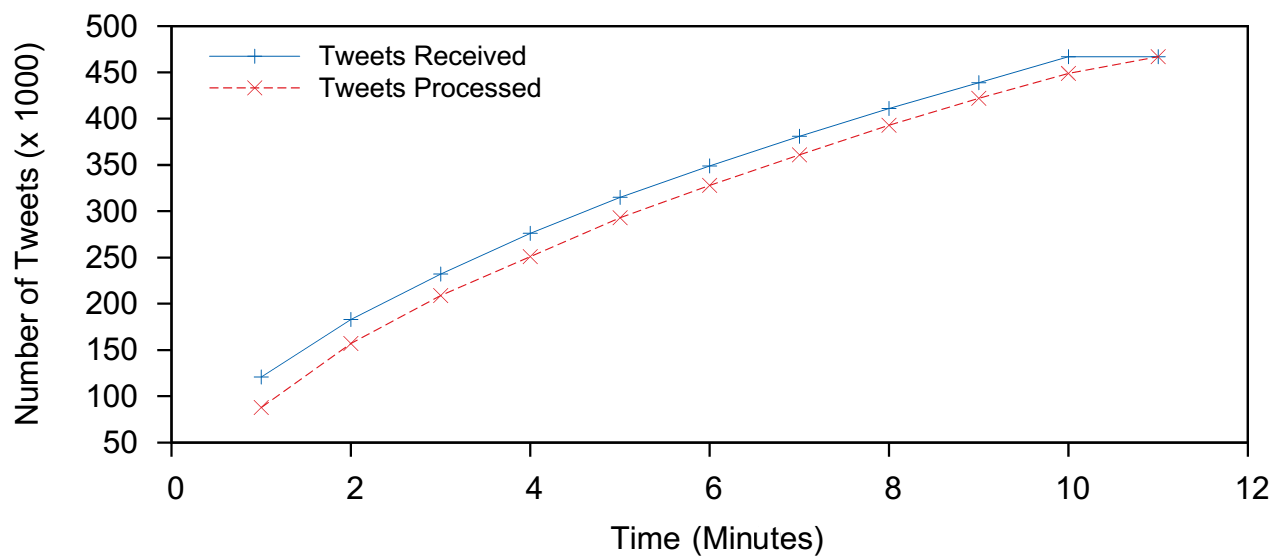}
  \caption{Number of received and process Tweets during the Experiment 3.} 
  \label{fig:exp3}
  \end{center}  
\end{figure}

\subsection{Experimental Summary}
Table \ref{table-summary} summarizes experimental results. The table shows a total number of processed Tweets, time took to process the Tweets in minutes, delay observed to process the Tweets in minutes, and speedup by comparing to Experiment 1. We consider a deployment with one virtual machine as a baseline and computed the speed up for Experiment 2 and 3. The larger size of cluster reducing latency and delay, however, the number of processed Tweets increases. The results indicate that increasing resources for the pipeline would provide processing in near to real-time. 
\begin{table}[h!]
\begin{center}
\caption{Experimental summary.}
\label{table-summary}
\begin{tabular}{ccccc}
 \hline
  \textbf{Exp\#} & \textbf{Tweets Processed} & \textbf{Processed Time (m)} & \textbf{Latency (m)} & \textbf{Speedup (\%)}\\
  \hline
  \rmfamily 1 & \rmfamily 464,200 & \rmfamily  15.0 & \rmfamily  5.0 & \rmfamily  - \\
  \rmfamily 2 & \rmfamily 468,600 & \rmfamily 11.5  & \rmfamily   1.5 & \rmfamily  130.0\\
  \rmfamily 3  &\rmfamily 466,700 &  \rmfamily 10.7  &  \rmfamily 0.7 & \rmfamily  140.0\\
  \hline
\end{tabular}
\end{center}
\end{table}

% Figure \ref{missing-requests} shows the number of missing Tweets per minutes during each experiment. The Tweets are missed mainly due to Apache Spark streaming module rejects the  does not write any data into Apcahe cassandra.
% \begin{figure}[!htb]
% \begin{center}
%   \includegraphics[width=\linewidth,height=5cm,keepaspectratio]{fig/summarygraph.png}
%   \caption{Number of Missing Request Graph}
%   \label{fig 3: Number of Missing Request Graph}
%   \end{center}
%   \label{missing-requests}
% \end{figure}
% \\The graph indicates that apache spark streaming application has large number of missing request per minutes when it run over the cluster size 1, less number of missing request as it run over larger cluster size. This indicates that as we cluster size increase number of missing request per minutes decrease which in turn increase overall processing of application.

%the graph in above paragraph can be rephrased as the the graph in figure 9 or something like that.
\section{Conclusion}
Real-time processing is involved in many applications to quickly respond the live events occurred in those applications. Twitter sentiment analysis is one of the most interesting techniques to find out the users opinion against a particular discussion, debate, or product. However performing such Twitter sentiment analysis against the large amount and high velocity of data requires large-scale processing of data on multiple machines using big data tools. The Real-time system we proposed and evaluated in this paper is able to perform the Twitter sentiment analysis over large and high velocity of data near the real-time. We conducted several experiments to determine the performance and scalability of the system against different cluster size and workload generator. Our proposed real-time processing system can also be used to develop real-time processing system for the other domains such as monitoring resources, fraud detection, and user stream analysis.

We have shown through experiments that the performance of the system is increased as we increase the cluster size. In future, we intend to improve the performance of our proposed system through enabling auto-scaling of appropriate component based on the workload. 

\bibliographystyle{IEEEtran}
\bibliography{refs.bib}
\end{document}